\documentclass[11pt]{article}

\usepackage{graphicx}
\usepackage{multirow}
\usepackage{amsmath,amssymb,amsfonts}
\usepackage{amsthm}
\usepackage{mathrsfs}
\usepackage[title]{appendix}
\usepackage{xcolor}
\usepackage{textcomp}
\usepackage{manyfoot}
\usepackage{booktabs}
\usepackage{algorithm}
\usepackage{algorithmicx}
\usepackage{algpseudocode}
\usepackage{listings}
\usepackage[colorlinks=true,linkcolor=blue,citecolor=blue,urlcolor=blue]{hyperref}
\usepackage[font=small,labelfont=bf]{caption}
\usepackage{enumitem}
\usepackage{changepage}

\theoremstyle{plain}

\theoremstyle{definition}

\theoremstyle{remark}

\newlength{\extralength}
\newcommand{\conflictsofinterest}[1]{\section*{Conflicts of Interest}#1}
\newcommand{\appendixtitles}[1]{}
\newcommand{\appendixstart}{}

\title{Mechanical Equilibrium in the Magnetized Quark--Hadron Mixed Phase: A Covariant Generalization of the Gibbs Condition}

\author{
Aric Hackebill\\
\\
\\
\texttt{aric.hackebill@uvm.edu}
}

\date{}

\begin{document}

\maketitle

\begin{abstract}
We formulate a covariant mechanical equilibrium condition for the quark-hadron mixed phase boundary in the presence of a magnetic-field-induced pressure anisotropy. Using the \emph{relativistic thin-shell} formalism to describe the quark-hadron boundary, we interpret conservation of stress-energy across the interface as a set of generalized Young--Laplace conditions which characterize the geometry of the interface. In a comoving stationary frame, this provides a covariant description of mechanical equilibrium at the interface, which serves as a replacement for the scalar pressure-balance condition used in the isotropic Gibbs construction.
\end{abstract}

\section{Introduction}

In the context of neutron stars (NS), the quark-hadron phase transition is typically studied in either the Maxwell or Gibbs construction \cite{glendenning_first-order_1992,schertler_neutron_1999,schertler_quark_2000,burgio_hadron-quark_2002,menezes_warm_2003,sharma_phase_2007,yang_influence_2008,alford_generic_2013,orsaria_quark_2014,xia_constraining_2019,ju_hadron-quark_2021,contrera_quark-nuclear_2022,wu_mixed_2025}, both of which assume isotropic pressures in each phase. In the Maxwell construction, both phases are taken to have equal pressures and baryonic chemical potentials, but distinct electronic chemical potentials arising from the requirement of independent charge neutrality in each phase. In the mixed phase of the Maxwell construction, the pressure remains constant, even as the baryon density increases from the pure hadron phase to the pure quark phase. This prohibits the mixed phase (as characterized by the Maxwell construction) from arising in NS since it would require a radially extended region in which the density increases without any corresponding increase in pressure. The Maxwell construction, therefore, allows for a sharp interface between the two phases over which the density jumps discontinuously, but it does not support a radially extended region occupied by both phases. 

In the Gibbs construction, the two phases are required to be globally charge neutral, and equilibrium of the electric chemical potentials is imposed along with pressure and baryonic chemical potential equilibrium between the two phases.
In contrast with the Maxwell construction, the Gibbs construction supports the possibility of a NS mixed phase since pressure varies with baryon density in that framework. We therefore restrict our attention to the Gibbs construction in order to study the quark-hadron mixed phase in NS.

Magnetic field effects on the quark–hadron mixed phase in the Gibbs construction have been studied under the assumption that the pressures arising in both phases are isotropic \cite{bandyopadhyay_quantizing_1997,rabhi_quarkhadron_2009,rather_magnetic-field_2023}. However, it was shown in \cite{ferrer_equation_2010,strickland_bulk_2012} that fermion systems immersed in a background magnetic field exhibit a pressure anisotropy with different longitudinal and transverse pressure components arising with respect to the local magnetic field direction. The effects of the pressure anisotropy on dense matter and neutron-star structure have been studied extensively \cite{isayev_anisotropic_2012,ferrer_thermodynamics_2019,ferrer_equation_2019,Chatterjee:2021wsr,bordbar_anisotropic_2022,Ferrer:2022wmy,ferrer_importance_2023,Peterson:2023bmr,Li:2024}. More
recently, magnetic-field-driven pressure anisotropies have also been investigated in the
context of neutron-star merger remnants \cite{Most:2025kqf}. Accordingly, the pressure equilibrium condition of the Gibbs construction must be modified to adequately describe the magnetized fermion sectors of the quark-hadron mixed phase. 

In order to develop the necessary modifications, we approximate the mixed phase as being divided into many Wigner–Seitz cells, each containing a volume fraction of both phases. Typically, the phase with the smaller volume fraction is treated as embedded within the other, with the embedded phase exhibiting a characteristic geometric structure. In the isotropic Gibbs construction, mechanical equilibrium at the interface is expressed by a single scalar pressure-balance condition, since the pressure in each phase is isotropic. As a result, the mechanical equilibrium condition does not depend on the interface geometry. In contrast, the different pressures of the anisotropic stress-energy tensor act differently on the interface depending on its orientation. Thus, the anisotropic case requires some specification of the interface geometry since the bulk pressures, on their own, do not fix the interface geometry.

The geometry of the mixed phase has previously been studied by minimizing the competition between surface and Coulomb energies to determine the preferred embedded structures, such as droplets, rods, slabs, tubes, and bubbles \cite{endo_charge_2006,endo_region_2011,yasutake_finite-size_2014,mariani_quark-hadron_2024,ju_effects_2025}. In this work, our goal is more limited. Rather than determining the preferred morphology or characteristic size of mixed phase structures, we develop the local mechanical equilibrium condition needed to describe anisotropic magnetized interfaces. In Sect. \ref{sec2} we use the \emph{relativistic thin-shell} formalism \cite{israel_singular_1966,schmidt_surface_1984,berezin_dynamics_1987,mansouri_equivalence_1996,poisson_relativists_2004} to model surface tension as stress-energy localized on the timelike worldtubes traced out by structures embedded in the mixed phase. The thin-shell formalism has been used elsewhere \cite{pereira_stability_2014,pereira_radial_2015} to study the stability of interfaces separating different layers (e.g., the core and crust) inside NS, but it has not been applied to mixed phase ``pasta" structures where surface stress-energy is used to model surface tension. Imposing stress-energy conservation across the phase boundary yields a set of covariant Young-Laplace conditions that relate the jump in the bulk stress-energy tensor to the geometry of the interface. The formalism, therefore, encodes both mechanical equilibrium and local shape information that any admissible mixed phase structure must satisfy.

The paper proceeds as follows. In Sect. \ref{sec2} we derive a general covariant interface stress-balance equation using thin-shell formalism. In Sect.~\ref{sec3}, we review the anisotropic bulk stress-energy tensor for magnetized fermion matter. In Sect.~\ref{sec4}, we derive the Young-Laplace-type stress-balance conditions for structures exhibiting constant surface tension and analyze embedded geometries that are compatible with those conditions. In Sect.~\ref{sec5}, we propose a minimal anisotropic surface stress-energy tensor model for magnetized systems and derive a corresponding pair of Young-Laplace stress-balance equations. In Sect.~\ref{sec6}, we use the mechanical equilibrium results of Sect. \ref{sec5} to derive a set of axisymmetric shape equations for droplet-type structures that are symmetric with the magnetic field direction. In Sect.~\ref{sec7}, we state the Gibbs conditions compatible with magnetized fermion systems. Finally, in Sect.~\ref{sec8}, we discuss how the covariant Young-Laplace formalism relates to studies of nuclear pasta geometries based on energy-minimization techniques.

\section{Covariant Mechanical Equilibrium Conditions}\label{sec2}

Let $\mathcal{S}$ denote the two-surface formed by the quark-hadron phase boundary of an embedded structure in the mixed phase. In general, the worldtube traced out by $\mathcal{S}$ is a three-dimensional hypersurface $\Sigma$ separating the spacetime manifold $\mathcal{M}$ into two bulk regions $\mathcal M_\pm$ occupied by distinct phases. The interface is assumed to be thin, so that physics on the boundary is described by a surface stress-energy tensor $S^{\mu\nu}$ \cite{israel_singular_1966}. 

The hypersurface $\Sigma$ can be defined as a level set $\phi(x)=0$ of a scalar function $\phi(x)$ chosen such that the spacelike normal to $\Sigma$ is

    \begin{equation}
        n_\mu := \frac{\nabla_\mu\phi}{\sqrt{\nabla_\alpha\phi\nabla^\alpha\phi}},\qquad n_\mu n^\mu=+1.
    \end{equation}

\noindent For simplicity, we assume $\phi(x)$ is chosen such that $\nabla_\alpha\phi\nabla^\alpha\phi=1$. 

In the thin-shell formalism, the stress-energy tensor takes the following distributional form

    \begin{equation}
        T^{\mu\nu}=T^{\mu\nu}_+\,\Theta(\phi)+T^{\mu\nu}_-\,\Theta(-\phi)+S^{\mu\nu}\delta(\phi),
    \end{equation}

\noindent where $T^{\mu\nu}_\pm$ are the bulk stress-energy tensors in $\mathcal M_\pm$ and $S^{\mu\nu}$ is the surface stress-energy tensor on $\Sigma$ (see \cite{poisson_relativists_2004}, Sect. 3.7.4). The unit normal is chosen such that it points from the ``$-$" to the ``$+$" sides of $\Sigma$. Imposing stress-energy conservation $\nabla_\nu{T^{\mu\nu}}=0$ yields a \emph{jump} condition

    \begin{equation}
        \label{eq:jump-balance-sec2}
        [T^{\mu\nu}]\,n_\nu=-\nabla_\nu S^{\mu\nu},
    \end{equation}

\noindent where $[T^{\mu\nu}]:=T^{\mu\nu}_+|_\Sigma-T^{\mu\nu}_-|_\Sigma$ represents the jump in bulk stress-energy across $\Sigma$. The jump condition may be written as a sum of components normal and tangential to $\Sigma$:

    \begin{equation}\label{eq:jump-balance-geom-sec2}
        [T^{\mu\nu}]\,n_\nu
        =
        \underbrace{n^\mu K_{\alpha\beta}S^{\alpha\beta}}_{\Sigma-normal}-\underbrace{h^\mu{}_\alpha D_\beta S^{\alpha\beta}}_{\Sigma-tangential}
        ,
    \end{equation}

\noindent where $h_{\mu\nu}:=g_{\mu\nu}-n_\mu n_\nu$ is the induced surface metric on $\Sigma$, $D_\beta$ is the surface covariant derivative, and $K_{\alpha\beta}$ is the extrinsic curvature of $\Sigma$ (see App. \ref{derivation} for details).

Eq. (\ref{eq:jump-balance-geom-sec2}) is the general covariant equilibrium condition on $\Sigma$ that relates the jump in bulk stress-energy (LHS) to the geometry of the interface (RHS). Its concrete physical content depends on the form of the bulk and surface stress-energy tensors. In the following sections, we specify $T^{\mu\nu}$ and $S^{\mu\nu}$ for anisotropic magnetized systems and show that Eq.~(\ref{eq:jump-balance-geom-sec2}) becomes a Young-Laplace-type mechanical equilibrium condition on the spatial phase-boundary $\mathcal{S}$ that determines allowed interface geometries.

\section{Anisotropic Stress-Energy Conditions}\label{sec3}

In \cite{ferrer_equation_2010}, the bulk stress-energy tensor for fermions immersed in a uniform background magnetic field was obtained through a semi-classical quantum-statistical average of the field-theoretic energy-momentum tensor. In a local Minkowski frame, the stress-energy tensor was found to have an explicit covariant form

    \begin{equation}
        T^{\mu\nu}
        =
        -\Omega\,\eta^{\mu\nu}
        +
        (\mu\rho+Ts)\,u^\mu u^\nu
        -
        BM\,\eta^{\mu\nu}_{\perp},
        \label{Tmunu}
    \end{equation}
    
\noindent where $\Omega=\Omega_f+\frac{B^2}{2}$ is the full thermodynamic potential, $\Omega_f$ is the fermionic thermodynamic potential, $\rho=-\frac{\partial\Omega}{\partial\mu}$ the number density, $s=-\frac{\partial{\Omega}}{\partial{T}}$ the entropy density, $M=-\frac{\partial\Omega}{\partial{B}}$ the magnetization, and $u^\mu$ the local fluid four velocity. The relative signs of terms proportional to $\eta^{\mu\nu}$ and $\eta^{\mu\nu}_\perp$ differ from the result found in \cite{ferrer_equation_2010} because that reference uses the $+,-,-,-$ convention, whereas we use the $-,+,+,+$ convention.  $\eta^{\mu\nu}_\perp$ is the projector onto the spatial directions perpendicular to the magnetic field given by

    \begin{equation}
        \eta^{\mu\nu}_\perp:=\Delta^{\mu\nu}-b^\mu{b}^\nu.
    \end{equation}

\noindent Here $\Delta^{\mu\nu}$ is the spatial projector orthogonal to $u^\mu$,

    \begin{equation}
        \Delta^{\mu\nu}:=\eta^{\mu\nu}+u^\mu u^\nu,
    \end{equation}
    
\noindent and $b^\mu$ is a spacelike unit four-vector specifying the magnetic-field direction

    \begin{equation}
        b^\mu:=\frac{B^\mu}{\sqrt{B_\mu{B}^\mu}},\quad\quad B^\mu:=\frac{1}{2}\epsilon^{\mu\nu\alpha\beta}{u_\nu}F_{\alpha\beta},
    \end{equation}

\noindent satisfying

    \begin{equation}
        u_\mu b^\mu=0,\qquad b_\mu b^\mu=1.
    \end{equation}

We work in a local approximation in which each Wigner-Seitz cell is taken to be sufficiently small that the magnetic field is approximately uniform throughout the cell. Accordingly, $\vec{B}$ and $b^\mu$ are treated as fixed background quantities shared by the coexisting phases within a given cell.    

It is convenient to reformulate Eq. (\ref{Tmunu}) as 

    \begin{equation}
        T^{\mu\nu}
        =
        \varepsilon\,u^\mu u^\nu
        +
        P_\perp\,\Delta^{\mu\nu}
        +
        (P_\parallel-P_\perp)\,b^\mu b^\nu,
        \label{Tmunuobvious}
    \end{equation}
    
\noindent where $\varepsilon$ is the energy density, and $P_\perp$ and $P_\parallel$ are the pressures perpendicular and parallel to the magnetic-field direction in a frame comoving with the fluid, i.e., $u^\mu=(1,0,0,0)$. In the $T\to0$ limit, these are given by: 

    \begin{align}\label{EOS}
        \varepsilon&=\Omega_f+\mu\rho+\frac{B^2}{2},
        \\
        P_\perp&=-\Omega_f-M_fB+\frac{B^2}{2},
        \\
        P_\parallel&=-\Omega_f-\frac{B^2}{2}.
    \end{align}

\noindent where $M_f=-\frac{\partial\Omega_f}{\partial{B}}$.

For a stationary interface in a local comoving Minkowski frame of the mixed phase, the world-tube $\Sigma$ is time-independent and the unit normal is purely spatial, $n^\mu=(0,\hat{\mathbf n})$, where $\hat{\textbf{n}}$ is the unit normal to the two-surface $\mathcal{S}$. In this case, the non-vanishing components of the LHS of Eq. (\ref{eq:jump-balance-geom-sec2}) are spatial and take the form

    \begin{equation}[T^{ij}]n_j=\Delta{P_\perp}n^i+\Delta{\Pi}b^i(\hat{\textbf{n}}\cdot\hat{\textbf{b}}),
    \end{equation}

\noindent where $\Delta\Pi:=\Delta(P_\parallel-P_\perp)$. This can be expressed in terms of components normal and tangential to $\mathcal{S}$ by inserting $b^i=(\hat{\textbf{n}}\cdot\hat{\textbf{b}})n^i+(b^i-(\hat{\textbf{n}}\cdot\hat{\textbf{b}})n^i)$, which gives

    \begin{equation}\label{LHS}             [T^{ij}]n_j=\underbrace{\left[\Delta{P_\perp}+\Delta\Pi(\hat{\textbf{n}}\cdot\hat{\textbf{b}})^2\right]n^i}_{\mathcal{S}-normal}+\underbrace{\Delta\Pi(\hat{\textbf{n}}\cdot\hat{\textbf{b}})\left[b^i-(\hat{\textbf{n}}\cdot\hat{\textbf{b}})n^i\right]}_{\mathcal{S}-tangential}.
    \end{equation}

\section{Constant Surface Tension and Admissible Geometries}\label{sec4}

The surface stress-energy tensor $S^{\mu\nu}$ can be derived from the surface action

    \begin{equation}\label{action}
        S_\Sigma=\int_\Sigma d^3x\,\sqrt{-h}\mathcal{L}_\Sigma
    \end{equation}

\noindent by varying with respect to the surface metric $h_{\mu\nu}$ 

    \begin{equation}\label{variation}
        S^{\mu\nu}\equiv \frac{-2}{\sqrt{-h}}\frac{\delta S_\Sigma}{\delta h_{\mu\nu}}.
    \end{equation}

\noindent In the case of constant surface tension, the surface energy is taken to be structureless and isotropically distributed over $\Sigma$, corresponding to $\mathcal{L}_\Sigma=\sigma$  \cite{lee_domain_2007}. Using $\delta \sqrt{-h}=\tfrac12 \sqrt{-h}\,h^{\mu\nu}\delta h_{\mu\nu}$ yields

    \begin{equation}\label{constant}
        S^{\mu\nu}=-\sigma h^{\mu\nu}.
    \end{equation}

\noindent Substituting this into Eq. (\ref{eq:jump-balance-geom-sec2}) gives

    \begin{equation}
        [T^{\mu\nu}]\,n_\nu =-\sigma{n}^\mu K_{\alpha\beta}\,h^{\alpha\beta},
    \end{equation}

\noindent where the induced metric compatibility condition $D_\rho{h}^{\mu\nu}=0$ was used. Thus the constant surface tension assumption leads directly to the vanishing of the tangential component of the stress balance equation. Shifting to a comoving frame and using $K_{ij}\,h^{ij}=\vec{\nabla}\cdot\hat{\textbf{n}}$ for $i,j=1,2,3$ (see Eq. (\ref{eq:K-spatial}) of App. \ref{derivation}) this becomes 

    \begin{equation}\label{temp}
        [T^{ij}]\,n_j =-\sigma(\vec{\nabla}\cdot\hat{\textbf{n}}){n}^i.
    \end{equation}

\noindent Substituting the magnetized jump condition, Eq. (\ref{LHS}), into Eq. (\ref{temp}), and contracting with $n_i$ and $h^j{}_i$ respectively, yields the normal and tangential components

    \begin{align}
        \left[\Delta{P_\perp}+\Delta\Pi(\hat{\textbf{n}}\cdot\hat{\textbf{b}})^2\right]&=-\sigma(\vec{\nabla}\cdot\hat{\textbf{n}}),\label{normalmain}
        \\
        \Delta\Pi(\hat{\textbf{n}}\cdot\hat{\textbf{b}})\left[b^j-(\hat{\textbf{n}}\cdot\hat{\textbf{b}})n^j\right] &=0.\label{tangentialmain}
    \end{align}

 Eq.~(\ref{normalmain}) relates the jump in bulk pressures across the interface to its curvature, while Eq.~(\ref{tangentialmain}) imposes a geometric restriction on admissible interface orientations. Taken together, Eqs.~(\ref{normalmain}) and (\ref{tangentialmain}) generalize the classical Young-Laplace condition describing the mechanical equilibrium of a capillary surface separating two liquid phases. 
 
 Interestingly, for a nonzero anisotropy jump $\Delta\Pi$, Eq.~(\ref{tangentialmain}) requires the surface normal $\hat{\mathbf n}$ to be either parallel, antiparallel, or orthogonal to the magnetic-field direction $\hat{\mathbf b}$ at every point on the interface. Constant surface tension, therefore, restricts the possible mixed-phase morphologies to the following classes:

\begin{description}[leftmargin=1cm,labelindent=1cm,labelsep=0.5cm,style=sameline]
    \item[\textbf{Slabs:}] extended interfaces for which $(\hat{\mathbf n}\cdot\hat{\mathbf b})=\pm1$ everywhere on $\mathcal S$;

    \item[\textbf{Generalized rods/tubes:}] interfaces satisfying $(\hat{\mathbf n}\cdot\hat{\mathbf b})=0$ everywhere on $\mathcal S$, corresponding to structures whose surfaces are generated by translating a fixed cross-sectional curve, in a plane orthogonal to $\hat{\textbf{b}}$, along $\hat{\textbf{b}}$.
\end{description}

Droplet-type geometries, by contrast, are not compatible with the constant-surface-tension ansatz, since their surface normals vary through a continuous range of angles relative to $\hat{\mathbf b}$ and therefore do not satisfy Eq.~(\ref{tangentialmain}) over the entire interface.

Importantly, Eq.~(\ref{normalmain}) also implies that the standard Gibbs pressure-equilibrium assumption, $P_+=P_-$, is incompatible with nonzero constant surface tension on curved interfaces, even in the isotropic case ($\vec{B}\to0$), since Eq. (\ref{normalmain}) would still require there to be a pressure jump across the interface proportional to its curvature.

\section{Anisotropic Surface Stress-Energy and the Associated Stress-Balance Conditions}\label{sec5}
The constant surface tension assumption considered in the previous section is too restrictive in the context of background magnetic fields. Just as the magnetic field generates an anisotropy in the bulk stress-energy tensor, it should also introduce a directional dependence in the surface stress-energy tensor. In principle, the surface stress-energy tensor should be derived from an underlying microphysical description, just as the bulk stress tensor is. Although some work has been done in this direction \cite{lugones_surface_2019}, here we adopt a minimal phenomenological model for the surface stress-energy tensor.

A covariant stress-energy tensor that is sensitive to the magnetic field direction can be obtained from the surface free-energy density ansatz $\mathcal{L}_\Sigma=\sigma(\eta)$ where $\eta=h^{\mu\nu}{b}_{\mu}b_{\nu}=1-(b\cdot{n})^2$ is a scalar encoding the tangential projection of the magnetic field onto $\Sigma$. This choice provides the minimal structure that is sensitive to the magnetic field direction while remaining invariant under $\hat{\textbf{n}}\to-\hat{\textbf{n}}$ (the latter is required because the sign of the interface normal is conventional).  Variation of the action, Eq. (\ref{action}), with respect to $h^{\mu\nu}$ gives

    \begin{equation}\label{anisoset}
        S^{\mu\nu}=-\sigma(\eta)h^{\mu\nu}+2\sigma'(\eta)q^\mu{q}^\nu,
    \end{equation}

\noindent where $q^\lambda=h^\lambda{}_\rho{b}^\rho$ is the tangential projection of $b^\rho$ onto $\Sigma$. 

Substituting Eq. (\ref{anisoset}) into the jump-balance condition, Eq. (\ref{eq:jump-balance-geom-sec2}), moving to comoving frame, and contracting with $n_i$ and $h^j{}_i$, respectively, yields the normal and tangential Young-Laplace conditions

    \begin{align}
        \left[\Delta{P_\perp}+\Delta\Pi(\hat{\textbf{n}}\cdot\hat{\textbf{b}})^2\right]&=\Big[-\sigma(\eta)\,\vec{\nabla}\cdot\hat{\textbf{n}}
        + 2\sigma'(\eta)\,\vec{q}\cdot((\vec{q}\cdot\vec\nabla)\hat{\textbf{n}})\Big],\label{normal3}
        \\
        \Delta\Pi(\hat{\textbf{n}}\cdot\hat{\textbf{b}})q^j&=2(\hat{\textbf{n}}\cdot\hat{\textbf{b}})\Big[\sigma'(\eta)\,\vec{\nabla}\cdot\hat{\textbf{n}}
        + 2\sigma''(\eta)\,\vec{q}\cdot((\vec{q}\cdot\vec\nabla)\hat{\textbf{n}})\Big] q^j.\label{tangential3}
    \end{align}

\noindent where $\vec{q}=\hat{\textbf{b}}-(\hat{\textbf{n}}\cdot\hat{\textbf{b}})\hat{\textbf{n}}$ (see App. \ref{derivation} for a derivation).

Eqs. (\ref{normal3}) and (\ref{tangential3}) describe the mechanical equilibrium conditions at the interface in the case of anisotropic surface stress-energy. Once again, both slabs orthogonal to $\hat{\textbf{b}}$ and generalized cylinders longitudinally aligned with $\hat{\textbf{b}}$ satisfy both balance equations. However, in contrast with the constant surface tension case, Eq. (\ref{tangential3}) allows for surfaces that are oblique to the field direction. In particular, droplet-type geometries are not ruled out by the absence of a tangential component in this case.

\section{Axisymmetric Shape Equations for Droplet-Type Geometries}
\label{sec6}

Let the magnetic field point along the $z$-axis, $\hat{\mathbf b}=\hat{\mathbf e}_z$, and consider an axisymmetric droplet described by the surface of rotation of the spherical radial graph

    \begin{equation}
        r=R(\theta),\qquad 0<\theta<\pi,\quad 0\le \varphi<2\pi.
    \end{equation}

The outward unit normal of the surface of rotation is computed in App. \ref{axisym} and may be expressed as

    \begin{equation}\label{eq:n-axisym}
        \hat{\mathbf n}(\theta)
        =
        \frac{R(\theta)\,\hat{\mathbf e}_r - R'(\theta)\,\hat{\mathbf e}_\theta}{\sqrt{R(\theta)^2+\big(R'(\theta)\big)^2}}.
    \end{equation}

\noindent Using $\hat{\mathbf e}_z=\cos\theta\,\hat{\mathbf e}_r-\sin\theta\,\hat{\mathbf e}_\theta$, we obtain

    \begin{equation}\label{eq:cosalpha}
        \hat{\mathbf n}\!\cdot\!\hat{\mathbf b}
        =
        \frac{R\cos\theta+R'\sin\theta}{\sqrt{R^2+(R')^2}}.
    \end{equation}

Eqs. (\ref{eq:n-axisym}) and (\ref{eq:cosalpha}) can be used to determine $\eta$, $\vec{\nabla}\cdot\hat{\textbf{n}}$, and $\vec{q}\cdot((\vec{q}\cdot\vec\nabla)\hat{\textbf{n}})$. In App. \ref{axisym} these are calculated to be

    \begin{align}
        \eta&=\frac{(R'\cos\theta - R\sin\theta)^2}{R^2+R'^2}\label{eta2},
        \\
        \vec{\nabla}\cdot\hat{\textbf{n}}&=\frac{
        2R^3
        - R^2R''
        - R^2R'\cot\theta
        + 3R(R')^2
        - (R')^3\cot\theta
        }{
        R\big(R^2+(R')^2\big)^{3/2}
        },\label{sphericalndot2}
        \\
        \vec{q}\cdot((\vec{q}\cdot\vec\nabla)\hat{\textbf{n}})&=\eta\frac{R^2 + 2(R')^2 - R R''}{(R^2+R'^2)^{3/2}}.
    \end{align}

\noindent Substituting these equations into Eqs. (\ref{normal3}) and (\ref{tangential3}) gives a set of axisymmetric shape equations for droplet-type interfaces.

\section{Anisotropic Mechanical Equilibrium in the Gibbs Construction with Surface Tension}
\label{sec7}

In the standard Gibbs construction chemical equilibrium implies a single baryon
chemical potential $\mu$ and a single electric chemical potential $\mu_e$,
which take the same values in both phases. Global electric charge neutrality in the Gibbs construction is imposed by

    \begin{equation}
        \chi\,\rho_Q^{\mathrm{HP}}+(1-\chi)\,\rho_Q^{\mathrm{QP}}=0,
        \label{eq:gibbs-neutrality}
    \end{equation}
    
\noindent where $0\le \chi \le 1$ is the hadron volume fraction and
$\rho_Q^{\mathrm{HP}}$ and $\rho_Q^{\mathrm{QP}}$ are the electric charge
densities of the hadronic and quark phases. Mechanical equilibrium is imposed by equating the bulk pressures in the two phases, $P_{HP}=P_{QP}$. Imposing both the chemical and mechanical equilibrium conditions then fixes the bulk parameters, e.g., $\mu$, $\mu_e$, etc., for a given value of $\chi$. 

The interface balance conditions derived in Sects.~\ref{sec4} and \ref{sec5}, however, show that in magnetized anisotropic systems this bulk pressure-equality condition must be replaced by the appropriate interface stress-balance conditions: Eqs.~(\ref{normalmain}) and (\ref{tangentialmain}) for isotropic surface tension, and Eqs.~(\ref{normal3}) and (\ref{tangential3}) for anisotropic surface stress-energy. Thus, in the presence of nonzero surface stress-energy, mechanical equilibrium is determined by the balance of bulk and surface stresses at the interface rather than by equality of bulk pressures alone. In particular, for curved interfaces, the pressures are generally discontinuous across the phase boundary. 

It is important to note that the modified equilibrium conditions are not purely thermodynamic relations since they now involve expressions encoding the geometry of the phase boundary. For instance, for droplet geometries Eqs. (\ref{normal3}) and (\ref{tangential3}) reduce to

    \begin{equation}\label{eq:pole-balance}
        \begin{aligned}
        \Delta P_\parallel
        &=
        -\sigma(0)\,\vec{\nabla}\cdot\hat{\mathbf n}\Big|_{\theta=0},
        \\
        &:=-\sigma(0){\kappa_\parallel},
        \end{aligned}
    \end{equation}

\noindent at the $\theta=0$ pole. Assuming that the bulk parameters, $\mu$ and $\mu_e$, are uniform in both phases, one might attempt to solve for them by imposing polar equilibrium, Eq. (\ref{eq:pole-balance}), along with charge conservation, Eq. (\ref{eq:gibbs-neutrality}), for each fixed value of $\chi$. However, these equations are not solvable unless a further constraint is introduced to determine the polar curvature $\kappa_\parallel$. This suggests that the modified Gibbs conditions require an additional geometric condition in order to construct the mixed phase self-consistently. Such a restriction likely arises as a size constraint on mixed-phase structures, which may be derived once Coulomb contributions are also considered.

As a result, the impact of these equilibrium conditions on the mixed-phase equation of state (EOS) cannot be determined from the modified mechanical equilibrium condition alone. In addition to specifying the bulk hadronic and quark models, a complete EOS construction would require supplementing the interface conditions with a size constraint, for example, by incorporating Coulomb and charge-screening effects. Since compact-star observables such as the mass-radius relation depend on the resulting EOS, their quantitative modification can only be determined after this self-consistent construction is carried out. Such a calculation is left for future work.

\section{Connection with the Energy-Minimization Formalism}\label{sec8}

It is useful to clarify how the covariant Young--Laplace condition derived above is related to the energy-minimization approach commonly used in Wigner--Seitz descriptions of the mixed phase. To this end, we compare the present formalism with the standard energy-minimization treatment of the compressible liquid-drop model (CLDM), following the presentation in \cite{ji_nuclear_2020}. One considers a Wigner-Seitz cell containing two phases separated by an interface. The free-energy density of the cell is given by

    \begin{equation}\label{freeenergy}
        f_{\rm cell}
        =
        \chi f_{-}(n_{-})
        +
        (1-\chi) f_{+}(n_+)
        +
        f_{\rm surf}(\chi,R,\sigma)
        +
        f_{\rm Coul}(\chi,R,n_+,n_-),
    \end{equation}
    
\noindent where $\chi$ is the volume fraction of the embedded phase, $\sigma$ is the surface tension, \(f_{-}\) and \(f_{+}\) are the bulk free-energy densities of the two phases, and \(f_{\rm surf}\) and \(f_{\rm Coul}\) denote the surface and Coulomb contributions. The bulk free-energies are taken to be dependent on the number densities  $n_+$ and $n_-$ associated with any number of particle species constituting each bulk phase. Finally, $R$ denotes the radius of a candidate embedded geometry of dimension $D$. The preferred configuration is then obtained by minimizing \(f_{\rm cell}\), subject to the relevant baryon-number and charge constraints, and comparing the minimized free energies for the different candidate geometries.

To compare this formalism with the Young--Laplace condition, we consider the isotropic \(B\rightarrow 0\) limit at zero temperature and omit the Coulomb contribution. Furthermore, to isolate the capillary contribution to the mechanical equilibrium condition, we also assume that the cell size is fixed so that $R$ and $\chi$ are linked. In this case, Eq. (\ref{freeenergy}) can be expressed directly in terms of energy densities 

     \begin{equation}
        \varepsilon_{\rm cell}
        =
        \chi \varepsilon_{-}(n_{-})
        +
        (1-\chi) \varepsilon_{+}(n_+)
        +
        \varepsilon_{\rm surf}(\chi,\sigma),
    \end{equation}

\noindent where the thermodynamic relation $F=E-TS$ was used. Imposing $\delta \varepsilon_{\rm cell}=0$ to determine the stationary value of $\varepsilon_{\rm cell}$ requires $\frac{\partial{\varepsilon_{\rm cell}(x_i)}}{\partial{x_i}}=0$ for each independent variable $x_i$ of $\varepsilon_{\rm cell}(x_i)$. In particular, the mechanical equilibrium condition $ \frac{\partial{\varepsilon_{\rm cell}}}{\partial\chi}=0$ gives

    \begin{equation}\label{minbalance}
        \Delta{P}=P_+-P_-=-\frac{\partial\varepsilon_{\rm surf}}{\partial{\chi}}
    \end{equation}

\noindent where $\varepsilon_i=-P_i+\sum_i{\mu_in_i}$ was used. 

For the standard Wigner--Seitz geometries, let
    \begin{equation}
        D=
        \begin{cases}
            3, & \text{spherical droplets or bubbles},\\
            2, & \text{cylindrical rods or tubes},\\
            1, & \text{planar slabs}.
        \end{cases}
    \end{equation}
    
\noindent The usual geometrical ansatz takes the cell to have the same symmetry as the embedded structure, so that

    \begin{equation}
        \chi=\left(\frac{R}{R_{\rm cell}}\right)^D,
    \end{equation}
    
\noindent where \(R\) is the radius, or half-thickness in the slab case, of the embedded structure. The surface-energy density is in each case

    \begin{equation} 
        \varepsilon_{\rm surf}
        =
        \frac{E_{\rm surf}}{V_{\rm cell}}
        =
        \frac{\sigma A_{\rm interface}}{V_{\rm cell}}
        =
        \frac{D\sigma \chi}{R}.
    \end{equation}
    
\noindent Using \(R=R_{\rm cell}\chi^{1/D}\), this can be written as

    \begin{equation}
        \varepsilon_{\rm surf}
        =
        \frac{D\sigma}{R_{\rm cell}}\chi^{(D-1)/D}.
    \end{equation}
    
\noindent Therefore

    \begin{equation}
        \frac{\partial \varepsilon_{\rm surf}}{\partial \chi}
        =
        \frac{(D-1)\sigma}{R}.
    \end{equation}

\noindent Substituting this into Eq.~\eqref{minbalance} yields

    \begin{equation}\label{eminbalance}
        \Delta{P}
        =
        -\frac{(D-1)\sigma}{R}.
    \end{equation}

The corresponding Young--Laplace conditions are given by Eqs. (\ref{normalmain}) and (\ref{tangentialmain}) in the $B\to0$ limit, which reduce to a single condition

    \begin{equation}
        \Delta{P}=-\sigma\vec{\nabla}\cdot\hat{\textbf{n}}.
    \end{equation}

\noindent Specializing to the same geometries considered above, we have for a sphere, cylinder, and slab, respectively

    \begin{equation}
        \vec{\nabla}\cdot \hat{\textbf{n}}
        =
        \begin{cases}
        2/R, & D=3,\\
        1/R, & D=2,\\
        0, & D=1,
        \end{cases}
    \end{equation}

\noindent or more compactly,

    \begin{equation}
        \vec{\nabla}\cdot \hat{\textbf{n}}
        =
        \frac{D-1}{R}.
    \end{equation}
    
\noindent The Young--Laplace condition becomes

    \begin{equation}
        \Delta{P}
        =
        -\frac{(D-1)\sigma}{R},
    \end{equation}
    
\noindent which agrees exactly with the result obtained through energy-minimization, Eq. (\ref{eminbalance}).

Thus, in the isotropic limit, the covariant Young--Laplace formalism reproduces the surface-tension part of the pressure-balance condition obtained via the CLDM-energy-minimization approach. When Coulomb effects are included, Eq.~\eqref{minbalance} is supplemented by the Coulomb contribution,

    \begin{equation}
        \Delta{P}
        =
        -\frac{\partial \varepsilon_{\rm surf}}{\partial \chi}
        -
        \frac{\partial \varepsilon_{\rm Coul}}{\partial \chi},
    \end{equation}
    
\noindent so that, in the isotropic case, the Young--Laplace term should be understood as the capillary contribution to the full pressure-balance condition.

 This clarifies what is new about the present work. When a magnetic field induces anisotropic bulk and/or surface stresses, the scalar pressure balance of the isotropic energy-minimization formalism should be replaced by the tensorial interface condition derived in Secs. \ref{sec4} and \ref{sec5}.

\section{Summary and Outlook}

In this work, we used the relativistic thin-shell formalism to describe the anisotropic mechanical equilibrium conditions at the interface separating relativistic magnetized quark and hadron phases. By modeling surface tension on the quark-hadron boundary as a thin shell and imposing stress-energy conservation across the interface, we arrived at generalized Young-Laplace conditions coupling the geometry of mixed phase structures with the jump in bulk stress-energy across the interface. The resulting formalism replaces the scalar pressure-balance condition $P_{HP}=P_{QP}$ conventionally used in the Gibbs construction. 
We then specialized this general framework to axisymmetric interfaces aligned with the magnetic field and derived a set of shape equations governing droplet geometries in the magnetized quark-hadron mixed phase. It should be emphasized that although the formalism was developed for anisotropic magnetized systems, it also shows that the standard equal-pressure condition is incompatible with a nonzero constant surface tension for curved structures, even in the absence of a magnetic field, since the interface curvature implies a corresponding jump in the bulk stress across the phase boundary. 

Looking forward, it remains to supplement the modified Gibbs conditions with a size constraint so that a full self-consistent construction of the quark-hadron mixed phase can be completed. One possibility is to include competing Coulomb effects, so that the characteristic scale emerges from the balance between surface and electrostatic contributions. Another option is to parameterize the curvature scale, for example through a prescription relating the curvature to the mixed-phase volume fraction $\chi$, and then solve simultaneously for the bulk thermodynamic variables and interface geometry. More broadly, large magnetic field-induced anisotropies may also destabilize some morphologies (e.g., droplets, slabs, rods, tubes, etc.) or modify the sequence of preferred structures relative to the isotropic case. Thus, a complete construction of the mixed phase likely requires a corresponding stability analysis of mixed phase morphologies in the presence of a strong magnetic field. Along this line, the axisymmetric droplet equations of Sect. \ref{sec6} may be probed to find stable droplet configurations. Finally, the form of the magnetized surface stress-energy tensor arising from a microphysical model requires further exploration. The framework developed in this paper provides the appropriate geometric and mechanical basis for pursuing these questions further.

\vspace{6pt}

\conflictsofinterest{The author declares no conflicts of interest.} 



\appendixtitles{yes}
\appendixstart
\appendix

\section{Derivation of the Generalized Young-Laplace Equations}\label{derivation}

Although relativistic thin-shell formalism and junction dynamics have been extensively studied elsewhere \cite{israel_singular_1966,schmidt_surface_1984,berezin_dynamics_1987,poisson_relativists_2004,mansouri_equivalence_1996}, in this appendix, we give a self-contained derivation of the covariant force-balance equations (generalized Young-Laplace equations) for a timelike worldtube interface separating two bulk phases. We work in four-dimensional spacetime with signature $(-,+,+,+)$ and use units $c=1$.

\subsection[levelset]{The Interface as a Level Set}

Let $\Sigma$ denote the three-dimensional timelike hypersurface traced out by the two-dimensional spatial interface $\mathcal{S}$ separating the bulk phases. We assume the spacetime $\mathcal{M}$ is divided by $\Sigma$ into two regions $\mathcal{M}_\pm$, and represent $\Sigma$ as the zero level set of a scalar function

    \begin{equation}
        \phi(x)=0,
    \end{equation}

\noindent such that the gradient $\nabla_{\mu}\phi$ in a neighborhood of $\Sigma$ is a unit spacelike covector, i.e.

    \begin{equation}
        \label{eq:signed-distance}
        \nabla_\mu \phi \,\nabla^\mu \phi = +1.
    \end{equation}

\noindent The unit normal to $\Sigma$ is then

    \begin{equation}
        \label{eq:normal-from-phi}
        n_\mu := \nabla_\mu \phi, 
        \qquad n^\mu n_\mu = +1.
    \end{equation}
    
\noindent We take $n_\mu$ to point from the ``$-$'' to ``$+$'' side of $\Sigma$. The projection tensor onto the tangent space of $\Sigma$, or the induced metric on $\Sigma$ (see \cite{carroll_spacetime_2019}, App. D), is 

    \begin{equation}
        \label{eq:gamma-projector}
        h_{\mu\nu} := g_{\mu\nu} - n_\mu n_\nu,
    \end{equation}

\noindent which satisfies

        \begin{align}
         h^\mu{}_\nu &= \delta^\mu{}_\nu - n^\mu n_\nu,
            \label{eq:proj-def}
              \\
             h^\mu{}_\nu n^\nu &= 0,
             \label{eq:proj-annihilates}
             \\
            h^\mu{}_\rho h^\rho{}_\nu &= h^\mu{}_\nu.
            \label{eq:proj-idempotent}
        \end{align}

\subsection{Distributional Stress-Energy Tensor with a Surface Layer}    

We assume that smooth bulk stress-energy tensors $T^{\mu\nu}_\pm$ are defined on $\mathcal{M}_\pm$, but that across $\Sigma$ the bulk stress may jump. This is typically modeled (see \cite{poisson_relativists_2004}, Sect. 3.7.4) as a thin shell carrying its own surface stress-energy tensor $S^{\mu\nu}$, supported on $\Sigma$, with distributional representation: 

    \begin{equation}
        \label{eq:dist-T}
        T^{\mu\nu}
        =
        T^{\mu\nu}_+ \,\Theta(\phi)
        +
        T^{\mu\nu}_- \,\Theta(-\phi)
        +
        S^{\mu\nu}\,\delta(\phi).
    \end{equation}

\noindent Physically, the surface layer carries stress and energy only within the surface and it does not transport stress-energy flux through the surface in the normal direction. Thus $S^{\mu\nu}$ is tangential in both indices:

    \begin{equation}
        \label{eq:S-tangent}
        S^{\mu\nu}n_\nu = 0,\quad\quad S^{\mu\nu}n_\mu=0.
    \end{equation}

\noindent The conservation of stress-energy is given by

    \begin{equation}\label{conservation}
        \nabla_\mu T^{\mu\nu}=0.
    \end{equation}

\subsection{Computing $\nabla_\nu T^{\mu\nu}$}
We now compute the covariant divergence of \eqref{eq:dist-T}. Using the distributional identities

    \begin{equation}
        \nabla_\nu \Theta(\phi) = \delta(\phi)\,\nabla_\nu \phi = \delta(\phi)\,n_\nu,
        \qquad
        \nabla_\nu \Theta(-\phi) = -\delta(\phi)\,n_\nu,
    \end{equation}
    
\noindent we obtain

    \begin{align}
        \nabla_\nu T^{\mu\nu}
        &=
        \nabla_\nu\!\left(T^{\mu\nu}_+ \Theta(\phi)\right)
        +
        \nabla_\nu\!\left(T^{\mu\nu}_- \Theta(-\phi)\right)
        +
        \nabla_\nu\!\left(S^{\mu\nu}\delta(\phi)\right)
        \nonumber\\[0.5em]
        &=
        (\nabla_\nu T^{\mu\nu}_+)\Theta(\phi)
        +
        T^{\mu\nu}_+\nabla_\nu\Theta(\phi)
        +
        (\nabla_\nu T^{\mu\nu}_-)\Theta(-\phi)
        +
        T^{\mu\nu}_-\nabla_\nu\Theta(-\phi)
        \nonumber\\
        &\hspace{2em}
        +
        (\nabla_\nu S^{\mu\nu})\delta(\phi)
        +
        S^{\mu\nu}\nabla_\nu\delta(\phi).
    \label{eq:divT-expand}
    \end{align}
    
\noindent Assuming the bulk tensors satisfy local conservation away from the interface,

    \begin{equation}
        \nabla_\nu T^{\mu\nu}_\pm = 0 \qquad \text{in }\mathcal{M}_\pm,
    \end{equation}
    
\noindent the $\Theta$-supported terms vanish. The remaining terms at $\phi=0$ become

    \begin{align}
        \nabla_\nu T^{\mu\nu}
        &=
        \delta(\phi)\,\bigl(T^{\mu\nu}_+ - T^{\mu\nu}_-\bigr)n_\nu
        +
        (\nabla_\nu S^{\mu\nu})\delta(\phi)
        +
        S^{\mu\nu}\nabla_\nu\delta(\phi)
        \nonumber\\
        &:=
        \delta(\phi)\,[T^{\mu\nu}]\,n_\nu
        +
        (\nabla_\nu S^{\mu\nu})\delta(\phi)
        +
        S^{\mu\nu}\delta'(\phi)\,\nabla_\nu\phi,
    \label{eq:divT-with-delta-prime}
    \end{align}
    
\noindent where in the last step we used the chain rule for distributions,

    \begin{equation}
        \nabla_\nu \delta(\phi)=\delta'(\phi)\,\nabla_\nu\phi=\delta'(\phi)\,n_\nu.
    \end{equation}
    
\noindent The $\delta'(\phi)$ term vanishes due to the tangentiality condition \eqref{eq:S-tangent}, leaving:

    \begin{equation}
        \label{eq:dist-balance}
        \nabla_\nu T^{\mu\nu}
        =
        \delta(\phi)\left( [T^{\mu\nu}]\,n_\nu + \nabla_\nu S^{\mu\nu}\right).
    \end{equation}
    
\noindent Imposing total stress-energy conservation (\ref{conservation}) in the distributional sense, $\nabla_\nu T^{\mu\nu}=0$, yields the surface force-balance law

    \begin{equation}
        \label{eq:jump-balance}
        [T^{\mu\nu}]\,n_\nu = - \nabla_\nu S^{\mu\nu}.
    \end{equation}

\subsection{$\nabla_\nu S^{\mu\nu}$ in Terms of Intrinsic Geometric Structures}
We now rewrite $\nabla_\nu S^{\mu\nu}$ in terms of intrinsic normal and tangential geometric structures on the hypersurface. Using (\ref{eq:proj-def}), the RHS of (\ref{eq:jump-balance}) becomes

    \begin{align}
        \nabla_\nu S^{\mu\nu}
        &=
        \nabla_\nu\left(\delta^\mu{}_\alpha\delta^\nu{}_\beta{S}^{\alpha\beta}\right)
        \\
        &=
        \nabla_\nu\!\left(h^\mu{}_\alpha h^\nu{}_\beta S^{\alpha\beta}\right)
        \nonumber\\
        &=
        (\nabla_\nu{h}^\mu{}_\alpha)h^\nu{}_\beta S^{\alpha\beta}
        +
        h^\mu{}_\alpha (\nabla_\nu{h}^\nu{}_\beta) S^{\alpha\beta}
        +
        h^\mu{}_\alpha h^\nu{}_\beta \nabla_\nu S^{\alpha\beta}.
    \label{eq:divS-step1}
    \end{align}
    
\noindent Combining the last two terms into a total derivative:

    \begin{equation}
        h^\mu{}_\alpha (\nabla_\nu{h}^\nu{}_\beta) S^{\alpha\beta}
        +
        h^\mu{}_\alpha h^\nu{}_\beta \nabla_\nu S^{\alpha\beta}
        =
        h^\mu{}_\alpha \nabla_\nu(h^\nu{}_\beta S^{\alpha\beta}).
    \end{equation}
    
\noindent Thus

    \begin{equation}
    \label{eq:divS-step2}
        \nabla_\nu S^{\mu\nu}
        =\underbrace{(\nabla_\nu{h}^\mu{}_\alpha)h^\nu{}_\beta S^{\alpha\beta}}_{\mathcal{A}^\mu}
        +
        \underbrace{h^\mu{}_\alpha \nabla_\nu(h^\nu{}_\beta S^{\alpha\beta})}_{\mathcal{B}^\mu}.
    \end{equation}
    
 \noindent We now compute $\mathcal{A}^\mu$ and $\mathcal{B}^\mu$ separately and then combine the results. 
 
 Using $h^{\mu}{}_{\alpha}=\delta^{\mu}{}_{\alpha}-n^\mu n_\alpha$ and the tangentiality condition $S^{\alpha\beta}n_\alpha=0$, we have

    \begin{align}
        \mathcal{A}^\mu
        &= \big[-(\nabla_\nu n^\mu)\,n_\alpha - n^\mu \nabla_\nu n_\alpha\big]\,
        h^{\nu}{}_{\beta}S^{\alpha\beta}
        \nonumber\\
        &= -n^\mu (\nabla_\nu n_\alpha)\,h^{\nu}{}_{\beta}S^{\alpha\beta},
        \nonumber\\
        &=-n^\mu (\nabla_\nu n_\alpha)\,h^{\nu}{}_{\beta}(\delta^\alpha{}_\sigma-n^\alpha{}n_\sigma)S^{\sigma\beta},
        \label{eq:A-step2}
        \nonumber\\
        &=-n^\mu (\nabla_\nu n_\alpha)\,h^{\nu}{}_{\beta}h^\alpha{}_\sigma{S}^{\sigma\beta}.
    \end{align}

\noindent Now we introduce the extrinsic curvature (see \cite{carroll_spacetime_2019}, App. D):

    \begin{equation}
        K_{\nu\alpha}=\nabla_\nu n_\alpha - n_\nu a_\alpha,
        \qquad
        a_\alpha:=n^\rho\nabla_\rho n_\alpha,
        \label{eq:K-def}
    \end{equation}
    
\noindent so that

    \begin{align}
        \mathcal{A}^\mu=&-n^\mu (K_{\nu\alpha}+n_\nu{a_\alpha})\,h^{\nu}{}_{\beta}h^\alpha{}_\sigma{S}^{\sigma\beta},\nonumber
        \\
        =&-n^\mu{h}^{\nu}{}_{\beta}h^\alpha{}_\sigma{K_{\nu\alpha}},{S}^{\sigma\beta}\nonumber
        \\
        =&-n^\mu{K_{\beta\sigma}}S^{\sigma\beta},\nonumber
        \\
        =&-n^\mu{K_{\alpha\beta}}S^{\alpha\beta},
        \label{Amu}
    \end{align}

\noindent where in the last step we have relabeled dummy indices and used the fact that the surface stress energy tensor is symmetric.

Next, we determine $\mathcal{B}^\mu$. Define

    \begin{equation}
        V^{\alpha\nu}:=h^{\nu}{}_{\beta}S^{\alpha\beta},
        \label{eq:V-def}
    \end{equation}

\noindent so that $n_\nu V^{\alpha\nu}=0$ (since $h^\nu{}_\beta n_\nu=0$). Then

    \begin{equation}
        \mathcal{B}^\mu=h^\mu{}_\alpha \nabla_\nu V^{\alpha\nu}.
        \label{eq:B-start}
    \end{equation}

\noindent Insert $\delta^\rho{}_\nu=h^\rho{}_\nu+n^\rho n_\nu$ on the contracted derivative index:

    \begin{align}
        \mathcal{B}^\mu
        &=h^\mu{}_\alpha(h^\rho{}_\nu+n^\rho n_\nu)\nabla_\rho V^{\alpha\nu}
        \nonumber\\
        &=h^\mu{}_\alpha{h}^\rho{}_\nu\nabla_\rho V^{\alpha\nu}
        +h^\mu{}_\alpha n^\rho n_\nu \nabla_\rho V^{\alpha\nu}.
        \label{eq:B-split}
    \end{align}

\noindent The first term is the intrinsic divergence:

    \begin{equation}
        h^\mu{}_{\alpha}h^\rho{}_\nu\nabla_\rho V^{\alpha\nu}
        =h^\mu{}_\alpha D_\beta S^{\alpha\beta},
        \label{eq:B-tan}
    \end{equation}

\noindent where $D_\beta$ is the intrinsic covariant derivative compatible with $h_{\mu\nu}$ (see \cite{carroll_spacetime_2019}, App. D). For the second term, differentiate $n_\nu V^{\alpha\nu}=0$:

    \begin{equation}
        0=\nabla_\rho(n_\nu V^{\alpha\nu})=(\nabla_\rho n_\nu)V^{\alpha\nu}+n_\nu\nabla_\rho V^{\alpha\nu},
    \end{equation}

\noindent hence $n_\nu\nabla_\rho V^{\alpha\nu}=-(\nabla_\rho n_\nu)V^{\alpha\nu}$. Therefore

    \begin{align}
        h^\mu{}_\alpha n^\rho n_\nu \nabla_\rho V^{\alpha\nu}
        &=-h^\mu{}_\alpha n^\rho(\nabla_\rho n_\nu)V^{\alpha\nu},
        \nonumber\\
        &=-h^\mu{}_\alpha n^\rho(\nabla_\rho \nabla_\nu\phi)V^{\alpha\nu},
        \nonumber\\
        &=-h^\mu{}_\alpha n^\rho(\nabla_\nu \nabla_\rho\phi)V^{\alpha\nu},
        \nonumber\\
        &=-h^\mu{}_\alpha n^\rho(\nabla_\nu n_\rho)V^{\alpha\nu},
        \nonumber\\
        &=0,
        \label{eq:B-normal-22}
    \end{align}

\noindent since $n^\rho(\nabla_\nu n_\rho)=\frac{1}{2}\nabla_\nu(n_\rho{n^\rho})=\frac{1}{2}\nabla_\nu(1)=0$. Putting (\ref{eq:B-tan}) and (\ref{eq:B-normal-22}) into (\ref{eq:B-split}) gives

    \begin{equation}
        \mathcal{B}^\mu=h^\mu{}_\alpha D_\beta S^{\alpha\beta}.
        \label{eq:B-final}
    \end{equation}

\noindent Finally, substituting Eqs.~\eqref{Amu} and \eqref{eq:B-final} into
Eq.~\eqref{eq:divS-step2} gives

    \begin{equation}\label{nablaSmunu}
        \nabla_\nu S^{\mu\nu}
        =-n^\mu K_{\alpha\beta}S^{\alpha\beta}+
        h^{\mu}{}_{\alpha}\,D_\beta S^{\alpha\beta}.
        \end{equation}

\subsection{Stationary World-Tube Viewpoint with Constant Surface Tension.}
Under the constant surface tension assumption $S^{\mu\nu}=-\sigma{h}^{\mu\nu}$ [Eq. (\ref{constant})], (\ref{nablaSmunu}) becomes 

        \begin{align}
            \nabla_\nu S^{\mu\nu}
            &={\sigma}n^\mu {K}_{\alpha\beta}h^{\alpha\beta},\nonumber
            \\
            &={\sigma}n^\mu(h^{\alpha\beta}\partial_\alpha n_\beta
        -h^{\alpha\beta}n_\alpha a_\beta),\nonumber
            \\
            &=\sigma{n^\mu}h^{\alpha\beta}\partial_\alpha n_\beta.
        \end{align}

 In a local comoving inertial frame, $\Sigma$ is time-independent and the unit normal to $\Sigma$ is purely spatial, $n^{\mu}=(0,n^i)$, so all normal geometry reduces to the Euclidean geometry of the spatial slices and only spatial components contribute. 

    \begin{align}
         \nabla_j S^{ij}&=\sigma{n^i}h^{ab}\partial_a n_b,
         \nonumber\\
         &=\sigma{n^i}(\delta^{ab}-n^a n^b)\,\partial_a n_b,
         \nonumber\\
         &=\sigma{n^i}\partial_a n^a,
         \nonumber\\
         &=\sigma{n^i}\vec{\nabla}\cdot\hat{\textbf{n}}.
        \label{eq:K-spatial}
    \end{align}

\subsection{Derivation of the Spatial Form of Eq.~(4) for the Anisotropic Surface Stress Tensor}

In the stationary comoving Minkowski approximation, the spacetime covariant derivative reduces to an ordinary derivative. Moreover, since the interface is stationary in the comoving frame, its normal has no temporal component, \(n^\mu=(0,n^i)\), and the jump condition has vanishing \(\mu=0\) component. Eq.~(\ref{eq:jump-balance-geom-sec2}) therefore reduces to a purely spatial condition on the interface two-surface

    \begin{equation}
        [T^{ij}]n_j
        =
        n^i K_{ab} S^{ab}
        -
        h^i{}_j D_k S^{jk},
        \label{eq:appendix_spatial_jump}
    \end{equation}

For the anisotropic surface stress tensor introduced in Eq.~(\ref{anisoset}), the spatial components are

    \begin{equation}
        S^{ij}=-\sigma(\eta)h^{ij}+2\sigma'(\eta)q^i q^j,
        \label{eq:appendix_Sij}
    \end{equation}

\noindent where $ q^i=h^i{}_j b^j=b^i-(\hat{\textbf{n}}\cdot \hat{\textbf{b}})n^i$ and $\eta=q_i q_i = 1-(\hat{\textbf{n}}\cdot \hat{\textbf{b}})^2$. We assume that each Wigner-Seitz cell is small enough that the magnetic field \(b^i\) in each cell is constant and uniform.

We now evaluate the normal and tangential pieces of Eq.~\eqref{eq:appendix_spatial_jump} separately. The normal part is given by

    \begin{align}
        n^iK_{ab}S^{ab}&=n^iS^{ab}\partial_a n_b,\nonumber
        \\
        &=[-\sigma(\eta)h^{jk}\partial_j n_k
        +
        2\sigma'(\eta)q^j q^k \partial_j n_k]n^i,\nonumber
        \\
        &= [-\sigma(\eta)\,\vec{\nabla}\cdot\hat{\textbf{n}}
        +
        2\sigma'(\eta)\,q^j q^k \partial_j n_k]n^i.
        \label{eq:appendix_normal_part}
    \end{align}
    
\noindent where in the last line Eq. (\ref{eq:K-spatial}) was used.

The tangential part is given by

    \begin{align}
        -h^i{}_j D_k S^{jk}
        &=
        -h^i{}_j h^\ell{}_k \partial_\ell S^{jk},\nonumber
        \\
        &=h^i{}_j h^\ell{}_k \partial_\ell\big(\sigma(\eta)h^{jk}\big)-
        2h^i{}_j h^\ell{}_k \partial_\ell \big(\sigma'(\eta) q^j q^k\big)\nonumber
        \\
        &=h^{i\ell}\partial_\ell \sigma(\eta)
        -
        2h^i{}_j h^\ell{}_k \partial_\ell \big(\sigma'(\eta) q^j q^k\big).
        \label{eq:appendix_tangent_start}
    \end{align}
    
\noindent where $h^i{}_{j}h^\ell{}_{k}\partial_\ell{h^{jk}}=0$ was used.

Because the magnetic field direction \(b^i\) is constant, all spatial variation arises through the normal \(n^i\). Let $c:=\hat{\textbf{n}}\cdot \hat{\textbf{b}}$. Then

    \begin{equation}
        q_i = b_i - c n_i, \qquad \eta = 1 - c^2 .
        \label{qeta}
    \end{equation}
    
\noindent Differentiating these expressions gives

    \begin{align}
        \partial_\ell c &= b_k \partial_\ell n_k = q_k \partial_\ell n_k, \label{1}\\
        \partial_\ell \eta &= -2c\,\partial_\ell c = -2c\,q_k \partial_\ell n_k, \label{2}\\
        \partial_\ell \sigma(\eta)
        &= \sigma'(\eta)\,\partial_\ell \eta
         = -2c\,\sigma'(\eta)\,q_k \partial_\ell n_k. \label{3}
    \end{align}
    
\noindent Hence

    \begin{align}
        h^{i\ell}\partial_\ell \sigma(\eta)
        &= -2c\,\sigma'(\eta)\,h^{i\ell} q^k \partial_\ell n_k, \notag\\
        &= -2c\,\sigma'(\eta)\,h^{i\ell} q^k \partial_k n_\ell, \notag\\
        &= -2c\,\sigma'(\eta)\,q^k h^i{}_j \partial_k n^j, \notag\\
        &= -2c\,\sigma'(\eta)\,q^\ell h^i{}_j \partial_\ell n^j ,
        \label{result2}
    \end{align}

\noindent where we used the level-set definition of $n^\mu$, which satisfies 
$\partial_\ell n_k = \partial_\ell \partial_k \phi = \partial_k \partial_\ell \phi = \partial_k n_\ell$.

Next, differentiating $q^j = b^j - c n^j$ yields

    \begin{align}
        \partial_\ell q^j &= -(\partial_\ell c)n^j - c\,\partial_\ell n^j, \label{11}\\
        h^i{}_j h^\ell{}_k \partial_\ell q^j &= -c\,h^\ell{}_k \partial_\ell n^i, \label{22}\\
        \partial_\ell\!\left(\sigma'(\eta) q^j q^k\right)
        &= \sigma''(\eta)(\partial_\ell \eta) q^j q^k
        + \sigma'(\eta)\Big[(\partial_\ell q^j)q^k + q^j(\partial_\ell q^k)\Big].
        \label{33}
    \end{align}

Substituting Eqs.~(\ref{result2}), (\ref{11}), and (\ref{33}) into Eq.~(\ref{eq:appendix_tangent_start}), and using
$h^i{}_j q^j = q^i$, $h^\ell{}_k q^k = q^\ell$, and
$h^\ell{}_k \partial_\ell n^k = \partial_k n_k$, we obtain

\begin{align}
-h^i{}_j D_k S^{jk}
&= -2c\,\sigma'(\eta)\,q^\ell h^i{}_j \partial_\ell n^j
   -2 h^i{}_j h^\ell{}_k \partial_\ell\!\left(\sigma'(\eta) q^j q^k\right) \notag\\
&= -2c\,\sigma'(\eta)\,q^\ell h^i{}_j \partial_\ell n^j
   -2\sigma''(\eta)(\partial_\ell \eta) q^i q^\ell
   -2\sigma'(\eta) q^k h^i{}_j h^\ell{}_k \partial_\ell q^j
   -2\sigma'(\eta) q^i h^\ell{}_k \partial_\ell q^k \notag\\
&= -2c\,\sigma'(\eta)\,q^\ell h^i{}_j \partial_\ell n^j
   +4c\,\sigma''(\eta)\,q^j q^k \partial_j n_k\,q^i
   +2c\,\sigma'(\eta)\,q^\ell h^i{}_j \partial_\ell n^j
   +2c\,\sigma'(\eta)\,\partial_k n_k\,q^i \notag\\
&= 2c\Big[\sigma'(\eta)\,\partial_k n_k
   + 2\sigma''(\eta)\,q^j q^k \partial_j n_k\Big] q^i .
\label{inter}
\end{align}

\noindent Combining Eqs.~(\ref{eq:appendix_normal_part}) and (\ref{inter}), the spatial jump condition becomes

    \begin{align}
        [T^{ij}]n_j=&
        \Big[-\sigma(\eta)\,\partial_k{n_k}
        + 2\sigma'(\eta)\,q^j q^k \partial_j n_k\Big] n^i
        + 2c\Big[\sigma'(\eta)\,\partial_k n_k
        + 2\sigma''(\eta)\,q^j q^k \partial_j n_k\Big] q^i,\nonumber
        \\
        =&\Big[-\sigma(\eta)\,\vec{\nabla}\cdot\hat{\textbf{n}}
        + 2\sigma'(\eta)\,\vec{q}\cdot((\vec{q}\cdot\vec\nabla)\hat{\textbf{n}})\Big] n^i\nonumber
        \\
        &+ 2c\Big[\sigma'(\eta)\,\vec{\nabla}\cdot\hat{\textbf{n}}
        + 2\sigma''(\eta)\,\vec{q}\cdot((\vec{q}\cdot\vec\nabla)\hat{\textbf{n}})\Big] q^i.\label{balance2}
    \end{align}

\setcounter{equation}{0}
\renewcommand{\theequation}{\thesection.\arabic{equation}}

\section{Axisymmetric Shape Equations}\label{axisym}
If the two-surface $\mathcal{S}$ is axisymmetric about the $z$-axis, it is useful to parametrize the surface in spherical coordinates

    \begin{equation}
        r=R(\theta),\qquad 0<\theta<\pi,\quad 0\le\varphi<2\pi,
    \end{equation}
    
\noindent and represent it as the level set $F(r,\theta)=0$ with

    \begin{equation}
        F(r,\theta):=r-R(\theta).
    \end{equation}

\noindent Since $F$ is axisymmetric, $\partial_\varphi F=0$, and the spherical gradient gives

    \begin{equation}
        \vec{\nabla} F=\hat{\mathbf e}_r\,\partial_r F+\frac{1}{r}\hat{\mathbf e}_\theta\,\partial_\theta F
        =\hat{\mathbf e}_r-\frac{R'(\theta)}{r}\,\hat{\mathbf e}_\theta.
    \end{equation}

\noindent The unit normal $\hat{\textbf{n}}$ pointing away from the surface can then be determined through 

    \begin{equation}\label{normal}
        \hat{\textbf{n}}(\theta)=\frac{\vec{\nabla} F}{\|\vec{\nabla} F\|}\Bigg\vert_{r=R(\theta)}
        =\frac{R(\theta)\,\hat{\mathbf e}_r-R'(\theta)\,\hat{\mathbf e}_\theta}{\sqrt{R(\theta)^2+(R'(\theta))^2}}.
    \end{equation}

\noindent Finally, the spherical divergence of $\hat{\textbf{n}}(\theta)$ is given by

    \begin{align}
        \vec{\nabla}\cdot \hat{\textbf{n}}
        &= \Bigg[\vec{\nabla}\cdot\frac{\vec{\nabla} F}{\|\vec{\nabla} F\|}\Bigg]\Bigg\vert_{r=R(\theta)}\nonumber
        \\
        &=\frac{
        2R^3
        - R^2R''
        - R^2R'\cot\theta
        + 3R(R')^2
        - (R')^3\cot\theta
        }{
        R\big(R^2+(R')^2\big)^{3/2}
        }.\label{sphericalndot}
    \end{align}

To specialize Eq.~(\ref{balance2}) to an axisymmetric surface about the magnetic-field direction
$\hat{\textbf{b}}=\hat {\textbf{e}}_z$, it is useful to introduce the shorthand

    \begin{equation}
        S(\theta) := \sqrt{R(\theta)^2 + (R'(\theta))^2},
        \qquad
        \hat{\textbf{t}}(\theta) := \frac{R'(\theta)\hat{\textbf{e}}_r + R(\theta)\hat{\textbf{e}}_\theta}{S(\theta)},\nonumber
    \end{equation}
    
\noindent where $\hat{\textbf{t}}$ is the unit tangent along $R(\theta)$. Using
$\hat{\textbf{e}}_z = \cos\theta\,\hat{\textbf{e}}_r - \sin\theta\,\hat{\textbf{e}}_\theta$ together with Eq.~(\ref{normal}), one finds

    \begin{align}
        c := \hat{\textbf{n}}\cdot \hat{\textbf{b}}\label{c}
        &= \frac{R\cos\theta + R'\sin\theta}{S}, \\
        \vec q :=\hat{\textbf{b}} - c\,\hat{\textbf{n}}\label{q}
        &= \frac{R'\cos\theta - R\sin\theta}{S}\,\hat t, \\
        \eta = 1-c^2 = q_i q_i
        &= \frac{(R'\cos\theta - R\sin\theta)^2}{S^2}\label{eta}.
    \end{align}

From Eq. (\ref{q}) and (\ref{eta}) we have that 

    \begin{equation}
        \vec{q}\cdot((\vec{q}\cdot\vec\nabla)\hat{\textbf{n}})=\eta\hat{\textbf{t}}\cdot((\hat{\textbf{t}}\cdot\vec\nabla)\hat{\textbf{n}}).
    \end{equation}

\noindent In general,

    \begin{equation}
        (\hat{\textbf{t}}\cdot\vec\nabla)\hat{\textbf{n}}=\frac{d\hat{\textbf{n}}}{d\ell}
    \end{equation}
    
\noindent since the directional derivative of a field in the direction of the unit tangent equals the derivative of that field with respect to the arc length along the corresponding curve. Since $d\ell=S{d}\theta$ we have 

    \begin{align}
        \vec{q}\cdot((\vec{q}\cdot\vec\nabla)\hat{\textbf{n}})=&\eta\hat{\textbf{t}}\cdot\Big(\frac{1}{S}\frac{d\hat{\textbf{n}}}{d\theta}\Big),\nonumber
        \\
        =&\eta\hat{\textbf{t}}\cdot\frac{1}{S}\frac{d}{d\theta}\!\left(\frac{R\hat{\textbf{e}}_r - R'\hat{\textbf{e}}_\theta}{S}\right),\nonumber
        \\
        =&\eta\hat{\textbf{t}}\cdot\Big(\frac{R^2 + 2(R')^2 - R R''}{S^3}\,\hat{\textbf{t}}\Big),\nonumber
        \\
        =&\eta\frac{R^2 + 2(R')^2 - R R''}{S^3}.\label{kappa}
    \end{align}

Substituting Eq.~(\ref{LHS}) into Eq.~(\ref{balance2}), and projecting with $n_i$ and $h^j{}_i$, respectively, yields normal and tangential components 

    \begin{align}
        \Delta P_\perp + \Delta\Pi\,c^2
        &= -\sigma(\eta)\,\vec{\nabla}\cdot\hat{\textbf{n}} + 2\sigma'(\eta)\,\vec{q}\cdot((\vec{q}\cdot\vec\nabla)\hat{\textbf{n}}),\label{normalapp}\\
        \Delta\Pi\,c{q^j}
        &= 2c\Big[\sigma'(\eta)\,\vec{\nabla}\cdot\hat{\textbf{n}} + 2\sigma''(\eta)\,\vec{q}\cdot((\vec{q}\cdot\vec\nabla)\hat{\textbf{n}})\Big]q^j.
        \label{tangentialapp}
    \end{align}

\noindent Finally, substituting Eqs.  (\ref{sphericalndot}), (\ref{c}), (\ref{eta}), and (\ref{kappa}) gives a set of axisymmetric shape equations in terms of $R''(\theta)$, $R'(\theta)$, $R(\theta)$, and $\theta$.



\begin{thebibliography}{999}

\bibitem{glendenning_first-order_1992}
Glendenning, N.K.
\newblock First-order phase transitions with more than one conserved charge:
  {Consequences} for neutron stars.
\newblock {\em Physical Review D} {\bf 1992}, {\em 46},~1274--1287.
\newblock {\url{https://doi.org/10.1103/PhysRevD.46.1274}}.

\bibitem{schertler_neutron_1999}
Schertler, K.; Leupold, S.; Schaffner-Bielich, J.
\newblock Neutron stars and quark phases in the {Nambu}--{Jona}-{Lasinio}
  model.
\newblock {\em Physical Review C} {\bf 1999}, {\em 60},~025801.
\newblock {\url{https://doi.org/10.1103/PhysRevC.60.025801}}.

\bibitem{schertler_quark_2000}
Schertler, K.; Greiner, C.; Schaffner-Bielich, J.; Thoma, M.H.
\newblock Quark phases in neutron stars and a third family of compact stars as
  signature for phase transitions1.
\newblock {\em Nuclear Physics A} {\bf 2000}, {\em 677},~463--490.
\newblock {\url{https://doi.org/10.1016/S0375-9474(00)00305-5}}.

\bibitem{burgio_hadron-quark_2002}
Burgio, G.F.; Baldo, M.; Sahu, P.K.; Schulze, H.J.
\newblock Hadron-quark phase transition in dense matter and neutron stars.
\newblock {\em Physical Review C} {\bf 2002}, {\em 66},~025802.
\newblock {\url{https://doi.org/10.1103/PhysRevC.66.025802}}.

\bibitem{menezes_warm_2003}
Menezes, D.P.; Providência, C.
\newblock Warm stellar matter with deconfinement: {Application} to compact
  stars.
\newblock {\em Physical Review C} {\bf 2003}, {\em 68},~035804.
\newblock {\url{https://doi.org/10.1103/PhysRevC.68.035804}}.

\bibitem{sharma_phase_2007}
Sharma, B.K.; Panda, P.K.; Patra, S.K.
\newblock Phase transition and properties of a compact star.
\newblock {\em Physical Review C} {\bf 2007}, {\em 75},~035808.
\newblock {\url{https://doi.org/10.1103/PhysRevC.75.035808}}.

\bibitem{yang_influence_2008}
Yang, F.; Shen, H.
\newblock Influence of the hadronic equation of state on the hadron-quark phase
  transition in neutron stars.
\newblock {\em Physical Review C} {\bf 2008}, {\em 77},~025801.
\newblock {\url{https://doi.org/10.1103/PhysRevC.77.025801}}.

\bibitem{alford_generic_2013}
Alford, M.G.; Han, S.; Prakash, M.
\newblock Generic conditions for stable hybrid stars.
\newblock {\em Physical Review D} {\bf 2013}, {\em 88},~083013.
\newblock {\url{https://doi.org/10.1103/PhysRevD.88.083013}}.

\bibitem{orsaria_quark_2014}
Orsaria, M.; Rodrigues, H.; Weber, F.; Contrera, G.A.
\newblock Quark deconfinement in high-mass neutron stars.
\newblock {\em Physical Review C} {\bf 2014}, {\em 89},~015806.
\newblock {\url{https://doi.org/10.1103/PhysRevC.89.015806}}.

\bibitem{xia_constraining_2019}
Xia, C.J.; Maruyama, T.; Yasutake, N.; Tatsumi, T.
\newblock Constraining quark-hadron interface tension in the multimessenger
  era.
\newblock {\em Physical Review D} {\bf 2019}, {\em 99},~103017.
\newblock {\url{https://doi.org/10.1103/PhysRevD.99.103017}}.

\bibitem{ju_hadron-quark_2021}
Ju, M.; Wu, X.; Ji, F.; Hu, J.; Shen, H.
\newblock Hadron-quark mixed phase in the quark-meson coupling model.
\newblock {\em Physical Review C} {\bf 2021}, {\em 103},~025809.
\newblock {\url{https://doi.org/10.1103/PhysRevC.103.025809}}.

\bibitem{contrera_quark-nuclear_2022}
Contrera, G.A.; Blaschke, D.; Carlomagno, J.P.; Grunfeld, A.G.; Liebing, S.
\newblock Quark-nuclear hybrid equation of state for neutron stars under modern
  observational constraints.
\newblock {\em Physical Review C} {\bf 2022}, {\em 105},~045808.
\newblock {\url{https://doi.org/10.1103/PhysRevC.105.045808}}.

\bibitem{wu_mixed_2025}
Wu, X.; Chu, P.C.; Ju, M.; Liu, H.
\newblock The mixed phase quark core in massive hybrid stars.
\newblock {\em The European Physical Journal C} {\bf 2025}, {\em 85},~362.
\newblock {\url{https://doi.org/10.1140/epjc/s10052-025-14088-y}}.

\bibitem{bandyopadhyay_quantizing_1997}
Bandyopadhyay, D.; Chakrabarty, S.; Pal, S.
\newblock Quantizing {Magnetic} {Field} and {Quark}-{Hadron} {Phase}
  {Transition} in a {Neutron} {Star}.
\newblock {\em Physical Review Letters} {\bf 1997}, {\em 79},~2176--2179.
\newblock {\url{https://doi.org/10.1103/PhysRevLett.79.2176}}.

\bibitem{rabhi_quarkhadron_2009}
Rabhi, A.; Pais, H.; Panda, P.K.; Providência, C.
\newblock Quark–hadron phase transition in a neutron star under strong
  magnetic fields.
\newblock {\em Journal of Physics G: Nuclear and Particle Physics} {\bf 2009},
  {\em 36},~115204.
\newblock {\url{https://doi.org/10.1088/0954-3899/36/11/115204}}.

\bibitem{rather_magnetic-field_2023}
Rather, I.A.; Rather, A.A.; Lopes, I.; Dexheimer, V.; Usmani, A.A.; Patra, S.K.
\newblock Magnetic-field {Induced} {Deformation} in {Hybrid} {Stars}.
\newblock {\em The Astrophysical Journal} {\bf 2023}, {\em 943},~52.
\newblock {\url{https://doi.org/10.3847/1538-4357/aca85c}}.

\bibitem{ferrer_equation_2010}
Ferrer, E.J.; de~la Incera, V.; Keith, J.P.; Portillo, I.; Springsteen, P.L.
\newblock Equation of state of a dense and magnetized fermion system.
\newblock {\em Physical Review C} {\bf 2010}, {\em 82},~065802.
\newblock {\url{https://doi.org/10.1103/PhysRevC.82.065802}}.

\bibitem{strickland_bulk_2012}
Strickland, M.; Dexheimer, V.; Menezes, D.P.
\newblock Bulk properties of a {Fermi} gas in a magnetic field.
\newblock {\em Physical Review D} {\bf 2012}, {\em 86},~125032.
\newblock {\url{https://doi.org/10.1103/PhysRevD.86.125032}}.

\bibitem{isayev_anisotropic_2012}
Isayev, A.A.; Yang, J.
\newblock Anisotropic pressure in dense neutron matter under the presence of a
  strong magnetic field.
\newblock {\em Physics Letters B} {\bf 2012}, {\em 707},~163--168.
\newblock {\url{https://doi.org/10.1016/j.physletb.2011.12.003}}.

\bibitem{ferrer_thermodynamics_2019}
Ferrer, E.J.; Hackebill, A.
\newblock Thermodynamics of neutrons in a magnetic field and its implications
  for neutron stars.
\newblock {\em Physical Review C} {\bf 2019}, {\em 99},~065803.
\newblock {\url{https://doi.org/10.1103/PhysRevC.99.065803}}.

\bibitem{ferrer_equation_2019}
Ferrer, E.J.; Hackebill, A.
\newblock Equation of {State} of a {Magnetized} {Dense} {Neutron} {System}.
\newblock {\em Universe} {\bf 2019}, {\em 5},~104.
\newblock {\url{https://doi.org/10.3390/universe5050104}}.

\bibitem{Chatterjee:2021wsr}
Chatterjee, D.; Novak, J.; Oertel, M.
Structure of ultra-magnetised neutron stars.
\textit{Eur. Phys. J. A} \textbf{2021}, \textit{57}, 249.
https://doi.org/10.1140/epja/s10050-021-00525-5.

\bibitem{bordbar_anisotropic_2022}
Bordbar, G.H.; Karami, M.
\newblock Anisotropic magnetized neutron star.
\newblock {\em The European Physical Journal C} {\bf 2022}, {\em 82},~74.
\newblock {\url{https://doi.org/10.1140/epjc/s10052-022-10038-0}}.

\bibitem{Ferrer:2022wmy}
Ferrer, E.J.; Hackebill, A.
Hadron-quark phase transition at finite density in the presence of a magnetic field: Anisotropic approach.
\textit{Int. J. Mod. Phys. A} \textbf{2022}, \textit{37}, 2250048.
https://doi.org/10.1142/S0217751X22500488.

\bibitem{ferrer_importance_2023}
Ferrer, E.J.; Hackebill, A.
\newblock The {Importance} of the {Pressure} {Anisotropy} {Induced} by {Strong}
  {Magnetic} {Fields} on {Neutron} {Star} {Physics}.
\newblock {\em Journal of Physics: Conference Series} {\bf 2023}, {\em
  2536},~012007.
\newblock {\url{https://doi.org/10.1088/1742-6596/2536/1/012007}}.

\bibitem{Peterson:2023bmr}
Peterson, J.; Costa, P.; Kumar, R.; Dexheimer, V.; Negreiros, R.; Provid\^encia, C.
Temperature and strong magnetic field effects in dense matter.
\textit{Phys. Rev. D} \textbf{2023}, \textit{108}, 063011.
https://doi.org/10.1103/PhysRevD.108.063011.

\bibitem{Li:2024}
Li, Y.H.; Ma, W.Q.; Wang, H.
Equation of state and anisotropy of pressure of magnetars.
\textit{Astronomische Nachrichten} \textbf{2024}, \textit{345}, e230180.
https://doi.org/10.1002/asna.20230180.

\bibitem{Most:2025kqf}
Most, E.R.; Peterson, J.; Scurto, L.; Pais, H.; Dexheimer, V.
Impact of Magnetic-field-driven Anisotropies on the Equation of State Probed in Neutron Star Mergers.
\textit{Astrophys. J. Lett.} \textbf{2025}, \textit{989}, L29.
https://doi.org/10.3847/2041-8213/adf62d.

\bibitem{endo_charge_2006}
Endo, T.; Maruyama, T.; Chiba, S.; Tatsumi, T.
\newblock Charge {Screening} {Effect} in the {Hadron}-{Quark} {Mixed} {Phase}.
\newblock {\em Progress of Theoretical Physics} {\bf 2006}, {\em
  115},~337--353.
\newblock {\url{https://doi.org/10.1143/PTP.115.337}}.

\bibitem{endo_region_2011}
Endo, T.
\newblock Region of a hadron-quark mixed phase in hybrid stars.
\newblock {\em Physical Review C} {\bf 2011}, {\em 83},~068801.
\newblock {\url{https://doi.org/10.1103/PhysRevC.83.068801}}.

\bibitem{yasutake_finite-size_2014}
Yasutake, N.; Łastowiecki, R.; Benić, S.; Blaschke, D.; Maruyama, T.;
  Tatsumi, T.
\newblock Finite-size effects at the hadron-quark transition and heavy hybrid
  stars.
\newblock {\em Physical Review C} {\bf 2014}, {\em 89},~065803.
\newblock {\url{https://doi.org/10.1103/PhysRevC.89.065803}}.

\bibitem{mariani_quark-hadron_2024}
Mariani, M.; Lugones, G.
\newblock Quark-hadron pasta phase in neutron stars: {The} role of
  medium-dependent surface and curvature tensions.
\newblock {\em Physical Review D} {\bf 2024}, {\em 109},~063022.
\newblock {\url{https://doi.org/10.1103/PhysRevD.109.063022}}.

\bibitem{ju_effects_2025}
Ju, M.; Wu, X.; Shen, H.
\newblock Effects of symmetry energy on the properties of a hadron-quark mixed
  phase in hybrid stars.
\newblock {\em Physical Review C} {\bf 2025}, {\em 111},~055801.
\newblock {\url{https://doi.org/10.1103/PhysRevC.111.055801}}.

\bibitem{israel_singular_1966}
Israel, W.
\newblock Singular hypersurfaces and thin shells in general relativity.
\newblock {\em Il Nuovo Cimento B (1965-1970)} {\bf 1966}, {\em 44},~1--14.
\newblock {\url{https://doi.org/10.1007/BF02710419}}.

\bibitem{schmidt_surface_1984}
Schmidt, H.J.
\newblock Surface layers in general relativity and their relation to surface
  tensions.
\newblock {\em General Relativity and Gravitation} {\bf 1984}, {\em
  16},~1053--1061.
\newblock {\url{https://doi.org/10.1007/BF00760644}}.

\bibitem{berezin_dynamics_1987}
Berezin, V.A.; Kuzmin, V.A.; Tkachev, I.I.
\newblock Dynamics of bubbles in general relativity.
\newblock {\em Physical Review D} {\bf 1987}, {\em 36},~2919--2944.
\newblock {\url{https://doi.org/10.1103/PhysRevD.36.2919}}.

\bibitem{mansouri_equivalence_1996}
Mansouri, R.; Khorrami, M.
\newblock The equivalence of {Darmois}‐{Israel} and distributional method for
  thin shells in general relativity.
\newblock {\em Journal of Mathematical Physics} {\bf 1996}, {\em
  37},~5672--5683.
\newblock {\url{https://doi.org/10.1063/1.531740}}.

\bibitem{poisson_relativists_2004}
Poisson, E.
\newblock {\em A {Relativist}'s {Toolkit}: {The} {Mathematics} of
  {Black}-{Hole} {Mechanics}}; Cambridge University Press: Cambridge,  2004.
\newblock {\url{https://doi.org/10.1017/CBO9780511606601}}.

\bibitem{pereira_stability_2014}
Pereira, J.P.; Coelho, J.G.; Rueda, J.A.
\newblock Stability of thin-shell interfaces inside compact stars.
\newblock {\em Physical Review D} {\bf 2014}, {\em 90},~123011.
\newblock {\url{https://doi.org/10.1103/PhysRevD.90.123011}}.

\bibitem{pereira_radial_2015}
Pereira, J.P.; Rueda, J.A.
\newblock {RADIAL} {STABILITY} {IN} {STRATIFIED} {STARS}.
\newblock {\em The Astrophysical Journal} {\bf 2015}, {\em 801},~19.
\newblock {\url{https://doi.org/10.1088/0004-637X/801/1/19}}.

\bibitem{lee_domain_2007}
Lee, B.H.; Lee, W.; Nam, S.; Park, C.
\newblock Domain wall cosmology and multiple accelerations.
\newblock {\em Physical Review D} {\bf 2007}, {\em 75},~103506.
\newblock {\url{https://doi.org/10.1103/PhysRevD.75.103506}}.

\bibitem{lugones_surface_2019}
Lugones, G.; Grunfeld, A.G.
\newblock Surface tension of hot and dense quark matter under strong magnetic
  fields.
\newblock {\em Physical Review C} {\bf 2019}, {\em 99},~035804.
\newblock {\url{https://doi.org/10.1103/PhysRevC.99.035804}}.

\bibitem{carroll_spacetime_2019}
Carroll, S.M.
\newblock {\em Spacetime and {Geometry}: {An} {Introduction} to {General}
  {Relativity}}; Cambridge University Press,  2019.
\newblock ISBN: 9781108770385, {\url{https://doi.org/10.1017/9781108770385}}.

\bibitem{ji_nuclear_2020}
Ji, F.; Hu, J.; Bao, S.; Shen, H.
Nuclear pasta in hot and dense matter and its influence on the equation of state for astrophysical simulations.
\textit{Physical Review C} \textbf{2020}, \textit{102}, 015806.
https://doi.org/10.1103/PhysRevC.102.015806.

\end{thebibliography}
\end{document}